\documentclass[journal,twoside,web]{ieeecolor}
\usepackage{generic}
\usepackage{cite}
\usepackage{amsmath,amssymb,amsfonts}
\usepackage{algorithmic}
\usepackage{microtype}
\usepackage{graphicx}
\usepackage{algorithm,algorithmic}
\usepackage{hyperref}
\hypersetup{hidelinks=true}
\usepackage{textcomp}
\usepackage{booktabs}
\usepackage{multirow}
\usepackage{bm}
\usepackage{soul}
\usepackage[table,xcdraw]{xcolor}
\def\BibTeX{{\rm B\kern-.05em{\sc i\kern-.025em b}\kern-.08em
    T\kern-.1667em\lower.7ex\hbox{E}\kern-.125emX}}
\begin{document}
\title{Integrating Bayesian Approaches and Expert Knowledge for Forecasting Continuous Glucose Monitoring Values in Type 2 Diabetes Mellitus}
\author{Yuyang Sun, \IEEEmembership{Student Member, IEEE}, and Panagiotis Kosmas., \IEEEmembership{Senior Member, IEEE}
}
\maketitle

\begin{abstract}
Precise and timely forecasting of blood glucose levels is essential for effective diabetes management. While extensive research has been conducted on Type 1 diabetes mellitus, Type 2 diabetes mellitus (T2DM) presents unique challenges due to its heterogeneity, underscoring the need for specialized blood glucose forecasting systems. This study introduces a novel blood glucose forecasting system, applied to a dataset of 100 patients from the ShanghaiT2DM study. Our study uniquely integrates knowledge-driven and data-driven approaches, leveraging expert knowledge to validate and interpret the relationships among diabetes-related variables and deploying the data-driven approach to provide accurate forecast blood glucose levels. The Bayesian network approach facilitates the analysis of dependencies among various diabetes-related variables, thus enabling the inference of continuous glucose monitoring (CGM) trajectories in similar individuals with T2DM. By incorporating past CGM data including inference CGM trajectories, dietary records, and individual-specific information, the Bayesian structural time series (BSTS) model effectively forecasts glucose levels across time intervals ranging from 15 to 60 minutes. Forecast results show a mean absolute error of \(6.41 \pm 0.60\) mg/dL, a root mean square error of \(8.29 \pm 0.95\) mg/dL, and a mean absolute percentage error of \(5.28 \pm 0.33\%\), for a 15-minute prediction horizon. This study makes the first application of the ShanghaiT2DM dataset for glucose level forecasting, considering the influences of diabetes-related variables. Its findings establish a foundational framework for developing personalized diabetes management strategies, potentially enhancing diabetes care through more accurate and timely interventions.
\end{abstract}

\begin{IEEEkeywords}
Bayesian network, Structure learning, Time series, Continuous glucose monitoring, Diabetes.
\end{IEEEkeywords}

\section{Introduction}
\label{sec:introduction}
Diabetes mellitus (DM) is a multifaceted metabolic disorder characterized by chronic hyperglycemia, posing a significant global health challenge. As estimated by the International Diabetes Federation, over 451 million individuals were diagnosed with diabetes globally in 2017, a number projected to rise to 693 million by 2045~\cite{cite_intro01}. The significant morbidity and mortality associated with DM, coupled with its complications, impose considerable strains on global healthcare infrastructures.

Continuous glucose monitoring (CGM) is a landmark technological development of diabetes management. By providing continuous, real-time glucose measurements, CGM plays a crucial role in daily DM management and in preventing related complications~\cite{cite_intro02}. Recent advancements in machine learning have inspired a growing interest in integrating machine learning techniques with CGM to enhance glucose level forecasting~\cite{cite_intro03,cite_intro04,cite_intro05,cite_intro06,cite_intro07}. These studies mainly focus on T1DM, which is characterized by the autoimmune destruction of pancreatic beta cells~\cite{cite_intro08}, using either OhioT1DM public dataset~\cite{cite_intro09} or data generated from the UVA/Padova simulator~\cite{cite_intro10}.

Conversely, Type 2 diabetes mellitus (T2DM) manifests primarily from insulin resistance and a gradual deterioration of beta-cell function~\cite{cite_intro08}. The inherent heterogeneity of T2DM, influenced by various pathogens, genetics, and lifestyle factors, demands an adapted approach in predictive modeling to capture these diverse influences. This study aims to fill this research gap by developing a CGM forecasting system for T2DM. Although T2DM accounts for approximately 90\% of all diabetes cases globally~\cite{cite_intro11}, the scarcity of available T2DM datasets has constrained research in this area. Our research uses the ShanghaiT2DM dataset~\cite{cite_intro12}, comprising data from 100 T2DM individuals. This dataset includes patient clinical information like anthropometric and biochemical characteristics, which is essential for the accurate CGM forecasting of T2DM by capturing its heterogeneity. It also contains time-series data of CGM measures and dietary intake records spanning 3 to 14 days. Additionally, a supplementary ShanghaiT1DM dataset, documenting data from 12 T1DM patients, serves for comparative analysis of the differences between T1DM and T2DM. To assist with understanding, Table~\ref{tab:my-table_0} lists acronyms and terms about diabetes-related characteristics used in this study.

\begin{table}[htb!]
\centering
\begin{tabular}{|l|l|}

\hline
\rowcolor[HTML]{C0C0C0} 
Characteristics                      & Acronyms \\ \hline
Body mass index                     & BMI      \\
Glycated haemoglobin                &HbA1c  \\
Glycated albumin                     & GA      \\
Total cholesterol                    & TC      \\
Triglyceride                         & TG      \\
High-density lipoprotein cholesterol & HDL-C   \\
Low-density lipoprotein cholesterol  & LDL-C    \\
Creatinine                           & CR      \\
Estimated glomerular filtration rate & eGFR    \\
Uric acid                            & UA      \\
Blood urea nitrogen                  & BUN     \\
Fasting plasma glucose              & FPG\\
2-hour postprandial plasma glucose  & 2HPP\\ 
\hline
\end{tabular}
\label{tab:my-table_0}
\caption{List of Acronyms for Diabetes-related Characteristics. }
\end{table}

\begin{figure*}[htb!]
  \centering
  \centerline{\includegraphics[width = 1\linewidth,height = 3.3in]{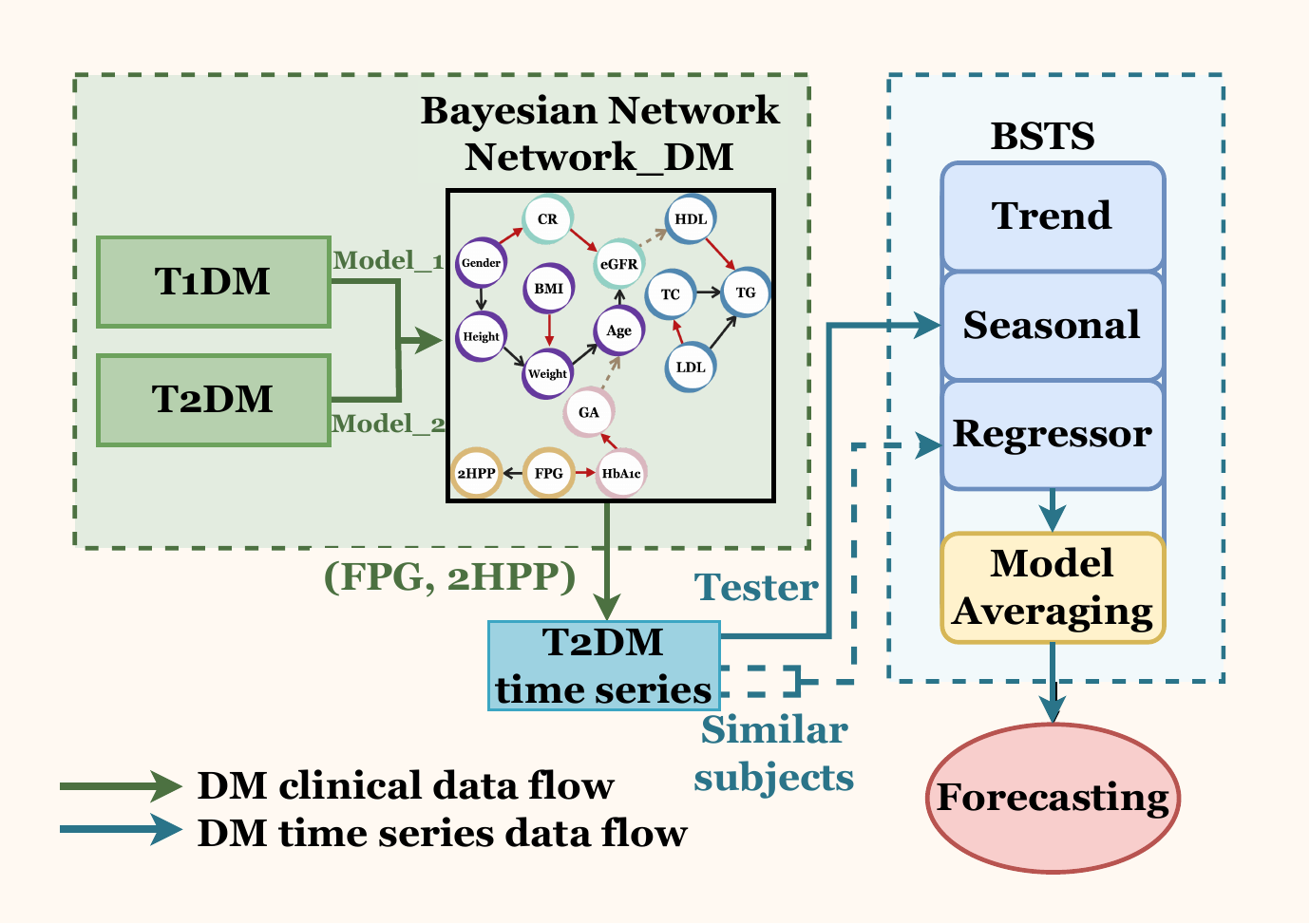}}
    \caption{Data-driven approach system architecture overview illustrating the integration of the Bayesian network and the BSTS models, accompanied by data flow representations. The 'DM clinical data flow' delineates the transmission of datasets encompassing anthropometric and biochemical characteristics related to both T1DM and T2DM. Concurrently, the 'DM time series data flow' illustrates the transmission of datasets that contain CGM measurements with dietary records. A detailed description of the Bayesian network is available in Section II-C and depicted in Figure~\ref{fig:fig_label_00}.}
\label{fig:fig_label_02}
\end{figure*}

Conventional data-driven approaches typically focus on training models for glucose level prediction across various predicted horizons (PHs). In contrast, our study integrates Bayesian networks generated through data-driven approaches with knowledge-driven analyses to validate and interpret complex interrelationships among diabetes-related variables. In our Bayesian network, identified arcs are categorized as causal, correlated, or independent, guided by expert knowledge. This approach enhances the interpretability of the Bayesian network, a crucial aspect that is frequently overlooked in biomedical research~\cite{cite_intro17,cite_intro13}. 

Structure learning, a key application of Bayesian networks, is pivotal in analyzing and inferring dependencies among variables. The foundational work by Pearl et al.~\cite{cite_dis02,cite_dis01} provides the basis for analyzing these dependencies. Its application facilitates discoveries across various domains, such as uncovering novel mechanisms in cellular systems \cite{cite_pre01} and interactions in psychiatric disorders \cite{cite_pre02}. Bayesian Networks offer an edge in interpretability over traditional black-box machine learning techniques by analyzing these dependencies, a requirement critically important in biomedical research. Dependencies in Bayesian Networks are sometimes interpreted as causal relationships~\cite{cite_intro17,cite_intro13}, but it is critical to differentiate between causality and conditional dependency, especially considering the impact of latent variables common in biomedical research. Therefore, our knowledge-driven analysis of the constructed Bayesian network, focuses on accurately identifying both these relationships between characteristics. For a deeper understanding of causality analysis, readers are referred to the provided references~\cite{cite_preadd01,cite_preadd02,cite_preadd03,cite_preadd04}.

In addition to Bayesian network structure learning, we deploy Bayesian structural time series (BSTS) modeling~\cite{cite_intro14,cite_intro15,cite_intro16} for accurate CGM forecasting. The large dataset of 100 T2DM individuals presents an opportunity to develop a new forecasting system that forecasts glucose levels and provides insights into the underlying mechanisms specific to T2DM. In contrast, the much smaller ShanghaiT1DM dataset mainly serves as a comparative reference in our study, enabling an in-depth analysis of the differences in glucose level forecasting variables between T1DM and T2DM by Beyesian network structure learning. A depiction of the data-driven approach architecture design employed for forecasting CGM values for T2DM is illustrated in Figure~\ref{fig:fig_label_02}. This forecasting system consists of two main models: a Bayesian Network and a BSTS model, which will be explained in detail in the Method section. The BSTS model generates the final predictions of our forecasting system, requiring previous CGM trajectories, daily records from the tester, and inferences from the Bayesian network as inputs for getting accurate forecasting.

\section{Method}
\subsection{Description of Datasets}
Our study uses the publicly available Shanghai DM datasets~\cite{cite_intro12}, which include detailed information about both T1DM and T2DM. Each DM dataset is classified into two key categories: time series data featuring CGM trajectories and dietary records, and individual clinical data detailing anthropometric and biochemical characteristics. These distinct data types are represented by the two separate data flows in Figure~\ref{fig:fig_label_02}.

\begin{table}[htb!]
\centering
\begin{tabular}{|l|l|l|}
\hline
\cellcolor[HTML]{C0C0C0}& \cellcolor[HTML]{C0C0C0}& \cellcolor[HTML]{C0C0C0}     \\  
\multirow{-2}{*}{\cellcolor[HTML]{C0C0C0}Data} & \multirow{-2}{*}{\cellcolor[HTML]{C0C0C0}
\begin{tabular}[c]{@{}l@{}}ShanghaiT1DM\\ (n=12)\end{tabular}} & \multirow{-2}{*}{\cellcolor[HTML]{C0C0C0}
\begin{tabular}[c]{@{}l@{}}ShanghaiT2DM\\ (n=100)\end{tabular}} \\ \hline
\multicolumn{3}{|l|}{\textbf{Anthropometric Characteristics}} \\
\hline
Age, years & $57.83\pm11.12$ & $60.17\pm 13.71$ \\
Female, n (\%) & $7 (58.3)$ & $44 (44.0)$ \\
Weight, kg & $57.39\pm10.18$ & $66.83\pm11.69$ \\
Height, m & $1.65\pm0.11$ & $1.66\pm0.09$ \\
BMI, kg/m\textsuperscript{2} & $21.08\pm4.05$ & $24.12\pm3.25$ \\
\hline
\multicolumn{3}{|l|}{\textbf{Biochemical Characteristics}} \\
\hline
HbA1c, mmol/mol & $89.38\pm36.63$ & $75.90\pm27.09$ \\
GA, \% & $32.25\pm15.97$ & $24.57\pm8.93$ \\
TC, mmol/L & $4.71\pm0.57$ & $4.87\pm1.16$ \\
TG, mmol/L & $1.07\pm0.52$ & $1.83\pm1.11$ \\
HDL, mmol/L & $1.42\pm0.52$ & $1.14\pm0.34$ \\
LDL, mmol/L & $2.82\pm0.76$ & $3.15\pm1.00$ \\
CR, umol/L & $53.78\pm18.27$ & $64.77\pm21.10$ \\
eGFR, ml/min/1.73\textsuperscript{2} & $142.73\pm62.15$ & $115.81\pm42.83$ \\
UA, mmol/L & $242.13\pm89.12$ & $335.20\pm95.63$ \\
BUN, mmol/L & $4.87\pm1.54$ & $6.13\pm1.99$ \\
\hline
\multicolumn{3}{|l|}{\textbf{Measurements}} \\
\hline
FPG, mg/dL & $193.2 (80.3,352.8)$ & $164.9 (55.8,432.0)$ \\
2HPP range, mg/dL & $268.8 (72.5,373.0)$ & $264.8 (97.0,610.4)$ \\
\hline
\multicolumn{3}{|l|}{\textbf{Time Series Characteristics}} \\
\hline
CGMs, mg/dL & $(39.6,424.8)$ & $(39.6,468.0)$ \\
Dietary intakes, unit/meal & $30.3\pm19.1$ & $17.4\pm10.5$ \\
\hline
\end{tabular}
\caption{Summary of anthropometric, biochemical characteristics, glucose measurements, and time series data extracted from the ShanghaiT1DM and ShanghaiT2DM datasets.}
\label{tab:my-table_1}
\end{table}

Specifically, the ShanghaiT1DM time series dataset includes records from $12$ T1DM patients, spanning periods ranging from $4$ to $14$ days. The ShanghaiT2DM time series dataset comprises data from $100$ T2DM patients, collected over periods extending from $3$ to $14$ days. These time series datasets include CGM measurements at $15$-minute intervals and dietary records logged three times daily. Notably, the raw dietary data, primarily consisting of food names and weights, was transformed for quantitative analysis. This involved calculating the glycemic load (GL) for each meal, providing a quantifiable assessment of its potential impact on blood glucose levels. The GL computation process is detailed in the data preprocessing subsection of the method section.

In addition to time series data, the ShanghaiDM datasets include records of each patient's anthropometric and biochemical characteristics, as listed in Table~\ref{tab:my-table_1}. In this table, $n$ represents the number of distinct subjects. Values related to anthropometric and biochemical characteristics are presented as mean $\pm$ standard deviation (SD) or as number (percentage). Data regarding the measurements, CGM values, and dietary intake records are exhibited in a range format (minimum, maximum) or as the mean glycemic load $\pm$ SD. While the original datasets contain additional features, including medical histories and complications, our study focuses on those quantifiable characteristics directly influencing blood glucose fluctuations. These characteristics include factors related to glucose metabolism, lipid profiling, kidney function, and dietary habits. Characteristics outside these domains were selectively excluded during data preprocessing for a more targeted analysis.

\subsection{Data Preprocessing}

As shown in Figure~\ref{fig:fig_label_02} and Table~\ref{tab:my-table_1}, the datasets are categorized into two groups: clinical data, which includes anthropometric and biochemical characteristics as well as glucose measurements used for Bayesian network construction, and time series data, which comprise CGMs and dietary intakes for BSTS forecasting. In the data preprocessing steps that follow, the first three methods are applied exclusively to clinical data to address missing values, noise, outliers, and non-standard data. The last method is reserved for processing time series data utilized in forecasting.

\textbf{Exclusion of Missing Values:} We removed samples with substantial missing clinical data to ensure data integrity. Specifically, $19$ samples from the ShanghaiT2DM dataset were excluded due to incomplete biochemical characteristics or missing FPG or 2HPP data. In the ShanghaiT1DM dataset, $1$ sample lacking 2HPP data was excluded. Additionally, UA and BUN characteristics were removed, as over 22\% of their values were missing. Unlike other glucose monitoring datasets \cite{cite_intro09, rodriguez2023t1diabetesgranada}, the ShanghaiT1DM and ShanghaiT2DM datasets contained no missing CGM data, eliminating the need for additional processing steps for missing values.

\textbf{Average Interpolation:} 
Following initial exclusions, clinical data in the ShanghaiT2DM dataset with minor missing characteristics (1 - 3 missing anthropometric or biochemical characteristics per sample) were interpolated with average values from corresponding characteristics in other samples. In this interpolation process, $32$ characteristics across $18$ samples were calculated and subsequently inserted into the missing positions. These interpolated characteristics constituted only 2\% of all clinical data, thereby minimally affecting the construction of the Bayesian network.

\textbf{Standardization and Encoding:}
After the exclusion and interpolation of missing values, the clinical data still varied in scale and unit, as listed in Table ~\ref{tab:my-table_1}. To reduce the effects of diverse magnitudes and variances on the construction of the Bayesian network and to cleanse potential noise and outliers, we introduced Z-score scaling. This method standardized the clinical data to have zero mean and a standard deviation of one. Subsequent to scaling, we applied encoding techniques to transform the standardized clinical data into discrete variables suitable for Bayesian network structure learning. Specifically, 'Gender' was encoded using one-hot encoding, which created distinct binary variables for each category. For other characteristics with multiple values, we utilized multi-class encoding, categorizing the data into four classes to more effectively capture the complexity of the dataset.

\textbf{Quantification:} Raw dietary records in the time series data, consisting of food items and weights, were quantitatively analyzed for their impact on blood glucose levels~\cite{cite_010}. We calculated the GL for each meal, a metric combining a food item's glycemic index (GI) and its available carbohydrate content ($CHO_{available}$), providing insight into how different foods affect blood glucose~\cite{cite_011,cite_012}:

\begin{align*}
    GL = \frac{GI \times CHO_{available}}{100} \tag{1}
\end{align*}

Here, $GL$ represents the glycemic load per 100 grams of food intake, $GI$ denotes the glycemic index of the food item, and $CHO_{available}$ is the available carbohydrate content in grams.

For multi-ingredient meals, GL values of individual food items were aggregated. GL values were primarily sourced from the University of Sydney's Human Nutrition Unit Database \cite{cite_005,cite_007,cite_008}. Additional sources, including mobile applications \cite{cite_003,cite_004,cite_009}, academic papers \cite{cite_002}, and books \cite{cite_006}, provided data for items not listed in the database, particularly those outside Western cuisine.

In cases where direct GL values were unavailable due to the diversity of dishes and cooking methods, we estimated GL based on similar food items. To ensure consistency in these estimations, we grouped similar dishes under common categories. For example, a dish of sliced pork with mushrooms and carrots served with rice was categorized as a pork and rice dish with an estimated $GI$ of $60$, $CHO_{available}$ of $23$, and a $GL$ of $13.8$. While these estimations are approximate, they are unlikely to significantly affect the overall results, as GL variations are mainly influenced by $CHO_{available}$ in staple foods like rice and noodles.

\subsection{Data-driven Bayesian Network Learning}
A Bayesian Network is a probabilistic graphical model, denoted as \( \mathcal{B} \), defined by a tuple \( (G, \bm{\Theta}) \), where \( G \) is a Directed Acyclic Graph (DAG) that represents dependencies among a set of random variables, and \( \bm{\Theta} \) includes parameters that define the strength of the conditional probability distributions of these dependencies governing the respective variables.

In this study, Bayesian networks are employed to learn structural dependencies among various anthropometric and biochemical characteristics. To optimize the network structure, we applied the Tabu search algorithm~\cite{cite_013}, a data-driven approach known for its efficiency and robustness in navigating complex, nonconvex solution spaces. This algorithm employs a greedy search strategy, systematically refining potential DAG structures, and uses the Bayesian information criterion (BIC) to evaluate and select the optimal model configuration. Tabu search enhances the learning process by avoiding revisiting previously explored configurations, thereby preventing cyclic dependencies and ensuring efficient exploration of new areas in the solution space. The BIC is instrumental in evaluating and selecting the model by penalizing excessive complexity, thus balancing the goodness of fit with the simplicity of the model.

Furthermore, to address the inherent challenges of nonconvex optimization and to ensure the robustness of the learned network, we incorporate Bootstrap re-sampling~\cite{cite_014,cite_015}. This technique is coupled with the Tabu search to generate multiple Bayesian networks from both the ShanghaiT1DM and ShanghaiT2DM datasets. During this process, each network's structure is constructed by assessing the frequency of arc occurrence across Bootstrap samples. Arcs that frequently appear across these samples are more reliable, indicating stronger and more consistent dependencies between variables. This frequency serves as an empirical measure of arc strength, with arcs exceeding a predefined strength threshold being selected for inclusion in the consensus network, 'Network\_DM'. The consensus network reduces the risk of overfitting to specific data idiosyncrasies and ensures that the final model reflects robust, statistically significant relationships.

The conditional probabilities are then calculated for the final structure during the parameter estimation phase, using maximum likelihood estimation methods, based on the arcs validated through the bootstrap Tabu search process. This approach ensures that the magnitudes of these probabilities accurately reflect the strength and nature of dependencies established by the structural learning phase, leading to a model that is both statistically rigorous and practically relevant. Subsequently, these coefficients are utilized to infer generated variables (FPG, 2HPP) based on the learned consensus network 'Network\_DM' for evaluating and selecting similar subjects, as shown in Figure \ref{fig:fig_label_02}.

After data preprocessing, data from $15$ ShanghaiT1DM and $91$ ShanghaiT2DM samples were utilized for constructing the 'Model\_1' and 'Model\_2' Bayesian networks, respectively. These networks were built through 100 iterative Bootstrap processes, enhancing their robustness. Arcs with a strength exceeding a threshold of $0.85$ were preserved, signifying their importance in representing underlying dependencies. The arc strength calculation is as follows:

\begin{align*}
    s(a_{ij}) = P(a_{ij})\tag{2}
\end{align*}

The strength of an arc \(a\) in a Bayesian network, represented as \( s(a_{ij}) \), is determined by \( P(a_{ij}) \), the probability of an arc's existence between nodes \( i \) and \( j \). This metric is used for identifying the existence of connections within the network, with the strength value indicating the likelihood of such connections.

\begin{figure*}[htb!]
  \centering
  \centerline{\includegraphics[width = 0.9\linewidth,height = 4.4in]{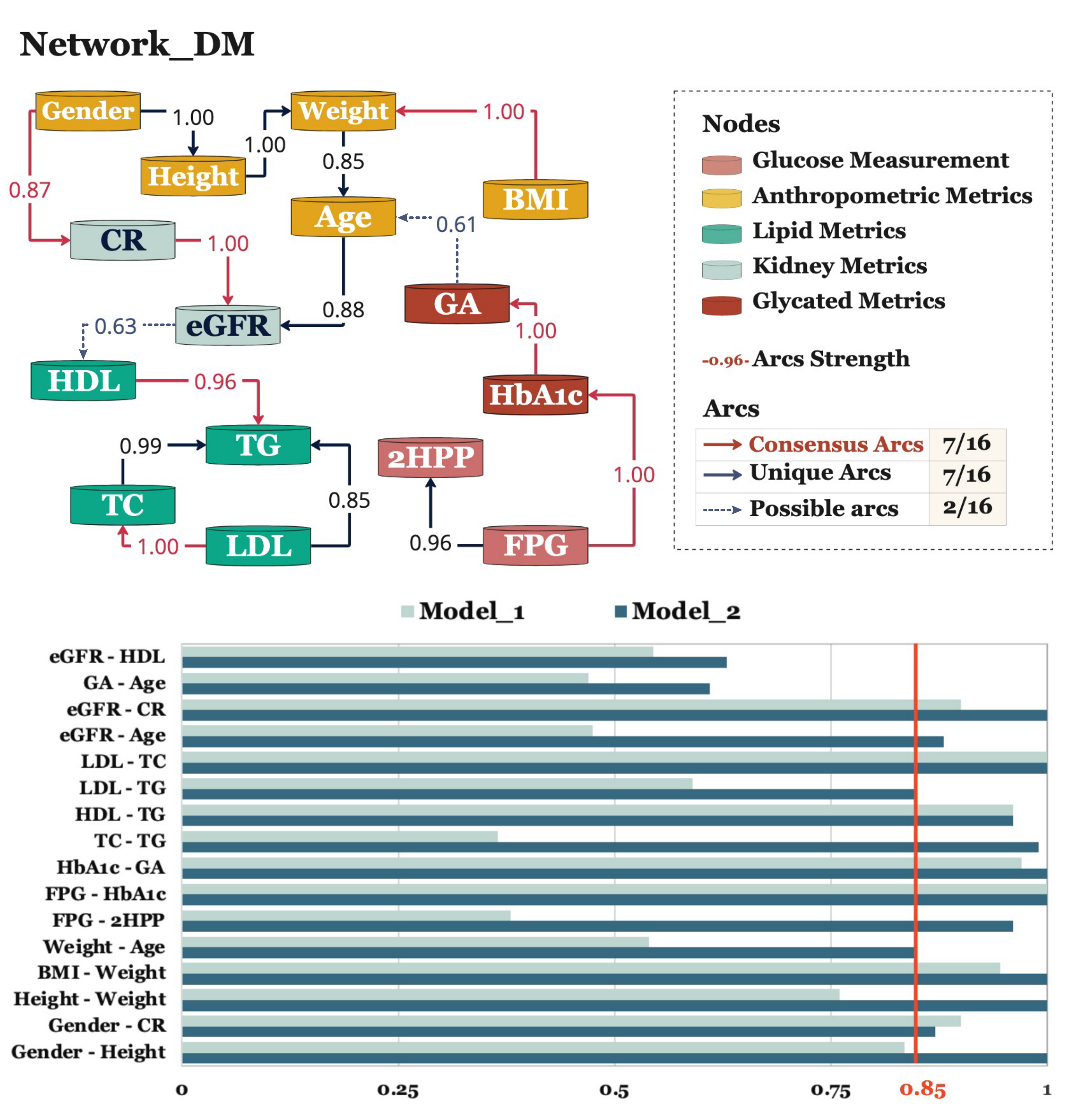}}
  \caption{The structure of 'Network\_DM' resulting from structural Bayesian network learning on both ShanghaiT1DM and ShanghaiT2DM datasets. This network depicts 16 arcs: 7 consensus arcs (red arrows), 7 unique arcs (black arrows), and 2 potential arcs (brown dotted arrows). 'Network\_DM' is an averaged representation based on $200$ structure learning models (both 'Model\_1' and 'Model\_2'), with each arc demonstrating a confidence level (arc strength) exceeding $0.85$. Nodes in the network correspond to characteristic features listed in Table \ref{tab:my-table_1}. Arc strengths for 'Network\_DM' are detailed at the bottom of the chart.}
\label{fig:fig_label_00}
\end{figure*}

The generated consensus Bayesian network, 'Network\_DM', is depicted in Figure~\ref{fig:fig_label_00}. Notably, 'Model\_1' lacks unique arcs compared to 'Model\_2', with all consensus arcs appearing in both models. Thus, the arcs displayed in 'Network\_DM' are the same as those existing in 'Model\_2'. The network's structure, as shown in Figure~\ref{fig:fig_label_00}, comprises several arcs and nodes representing glucose measurements and other characteristics. These nodes are divided into four metrics and one measurement for knowledge-driven analysis, as detailed in Section IV-C.

The 'arc strength' of 'Network\_DM' is visually depicted at the bottom of Figure~\ref{fig:fig_label_00}. The y-axis lists all arcs within 'Network\_DM', while the x-axis denotes their respective arc strengths.  Within the structure learning of Bayesian networks, 'arc strength' quantitatively denotes the probability of a potential edge (arc) between two nodes, considering its presence and direction. Essentially, it serves as a measure of confidence in the existence and directionality of probability dependencies between variables, ranging between 0 (low probability) to 1 (high probability).

All arcs exceeding the preset threshold in arc strength will be retained in the Bayesian network 'Network\_DM'. As observed in Figure~\ref{fig:fig_label_00}, every consensus and unique arc within 'Network\_DM' has arc strength exceeding the threshold of $0.85$ with a determined direction. The consensus arcs, marked by red arrows, demonstrate dependencies common to both the ShanghaiT1DM and ShanghaiT2DM datasets. Meanwhile, unique arcs, symbolized by black arrows, indicate dependencies unique to the ShanghaiT2DM dataset. Differences in arcs between the two networks are due to the sample size differences, with the ShanghaiT1DM dataset's limited size constraining its ability to identify potential arcs, especially at higher confidence thresholds. The possible arcs (brown dotted arrows) do not meet the strength threshold but are the strongest of all remaining arcs which can ensure a full connection of all available nodes to maximize the consideration of all variables. 

\begin{algorithm}
\caption{Selection of Similar Individuals}
\begin{algorithmic}
\STATE \textbf{Input:} Predict FPG and 2HPP results $P_{FPG}$ and $P_{2HPP}$ of $N$ individuals, Tester's $FPG$ and $2HPP$ values, Number of similar individuals $M$.
\STATE \textbf{Output:} Set $S$ of $M$ similar individuals.
    \FOR{$j = 1$ to $N-1$}
        \STATE Calculate Euclidean Distance $ED_j$:
        \STATE $ED_{j} \gets \sqrt{(P_{FPG_{j}}-FPG)^2 + (P_{2HPP_{j}}-2HPP)^2}$
        \ENDFOR
\FOR{$i = 1$ to $M$}
        \STATE  $S$ $\gets$ individual $j$ with smallest $ED_{j}$
        \STATE Remove $ED_{j}$ from $ED$
\ENDFOR
\RETURN $S$
\end{algorithmic}
\end{algorithm}

Following the construction of the Bayesian network, the individual's anthropometric and biochemical data are inputted to predict corresponding 'FPG' and '2HPP' values. These inferred values, treated as generated variables, provide a metric for assessing similarity degrees by considering all correlated variables and individual differences. They serve as metrics for calculating similarity, which are used for selecting similar distributions of selected testers during the BSTS forecasting process. Compared to a single variable, these values more accurately reflect an individual's metabolic similarity, encapsulating a broad spectrum of diabetes-related characteristics. To identify similar distributions, we calculate Euclidean distances from each individual's inferred 'FPG' and '2HPP' to the tester's original values. The two distributions exhibiting the shortest distances are then selected as the most similar. These selected distributions are subsequently incorporated into the BSTS forecasting phase, significantly enhancing the model’s performance in predicting time-series CGM values. The process of determining similarity among individuals is detailed in Algorithm 1.

\subsection{Knowledge-driven of Bayesian Network Structure Learning}

We analyze and categorize arcs within the Bayesian network, crucial for validating its structure and providing an analytical explanation of relationships between variables. Our aim is not just to categorize these arcs but to validate the network's structure through integrating data-driven and knowledge-driven approaches. The data-driven approach is generated by the Bayesian network learning, the knowledge-driven approach draws upon expert knowledge. These two driven approaches integration are inspired by Sachs et al.~\cite{cite_pre01}, offering a comprehensive method for recognizing reasonable relationships among diabetes-related variables.

In our previous classification, arcs were classified as consensus or unique arc types based on their presence in both 'Model\_1' and 'Model\_2'or not. However, a deeper analysis involves categorizing these arcs through expert knowledge, extensive interventional experiments, and robust statistical evidence, as causal, correlated, or independent. This categorization is crucial for two reasons: it provides interpretability of the network by elucidating the nature of the dependencies among variables, and it serves as a knowledge-driven validation of the Bayesian network's structure. Therefore, we rely on expert knowledge from established literature, adopting the methodology employed by Sachs et al.~\cite{cite_pre01}, to analyze the arcs in the Bayesian network 'Network\_DM' and to infer causal, correlated, or independent relationships between variables. 

For a clearer understanding, we have classified the arcs in the Bayesian network into three categories, and Table~\ref{tab:my-table_6} thoroughly enumerates all the arcs, organized based on our literature survey.

\textbf{Causal Relationships:} Relationships in this category are supported by well-established research and validated as causal through comprehensive interventional studies.

\textbf{Correlated Relationships:} Relationships in this category are recognized as correlations in the existing literature. However, their classification as causal remains uncertain. Further investigation is required to determine if latent common variables influence these nodes.

\textbf{Independent Relationships:} In the absence of evidence for either causal or correlated relationships in existing literature, relationships corresponding to possible arc types are categorized as independent. 

For the sake of brevity, the following subsections will focus on discussing the arcs that are categorized as causal relationships.

\begin{table}
    \centering
    \begin{tabular}{ccccc}
    \toprule
        From & To & Type& Category&References\\ \hline
        FPG& HbA1c&C& Correlated&\cite{cite_016,cite_017,cite_dis03,cite_018,cite_019,cite_020,cite_021} \\
        HbA1c&GA&C& Correlated&\cite{cite_022,cite_023,cite_024} \\
        FPG&2HPP&U& Correlated&\cite{cite_intro08,cite_025,cite_dis04,cite_dis05}  \\
        Gender & Height&U & Causal&\cite{cite_026}\\
       BMI  & Weight& C&Correlated&\cite{cite_027} \\
       Height  & Weight&U& Correlated&\cite{cite_027} \\
       Weight  & Age&U &Correlated&\cite{cite_028,cite_029,cite_030,cite_031}\\
        TC&TG&U &Correlated&\cite{cite_034,cite_035,cite_036,cite_039}\\
         LDL-C&TG&U&Correlated&\cite{cite_034,cite_035,cite_036,cite_038,cite_039} \\
         LDL-C&TC&C &Correlated&\cite{cite_036,cite_038,cite_039}\\
         HDL-C&TG&C &Correlated&\cite{cite_034,cite_035,cite_036,cite_037,cite_039}\\
       Age  & eGFR&U &Causal&\cite{cite_042,cite_043,cite_dis10} \\
       Gender  & CR&C & Causal&\cite{cite_042,cite_043,cite_dis11}\\
        CR & eGFR& C & Causal&\cite{cite_042,cite_043,cite_dis09}\\
         GA& Age &P& Independent& None\\
         eGFR& HDL-C&P& Independent& None\\ \bottomrule
    \end{tabular}
    \caption{Summary of the Bayesian network arcs types, categories, and supported literature. Type refers to arc presence situation ( C, Consensus; U, Unique; P, Possible).}
    \label{tab:my-table_6}
\end{table}

\begin{itemize}
    \item \textbf{Causal Relationships in the Bayesian Network}
\end{itemize}

The first causal arc, from 'Gender' to 'Height', reflects the significant influence of gender on height. Supported by global statistics and research~\cite{cite_026}, this relationship is generally attributed to genetic and hormonal factors, with males often being taller than females on average.

Other three causal arcs involve human kidney metrics: 'CR' to 'eGFR', 'Gender' to 'CR', and 'Age' to 'eGFR'.CR is a waste product of creatine breakdown in muscles, filtered and excreted by the kidneys. eGFR measures the rate at which kidneys filter blood. Both CR and eGFR are crucial indicators of kidney function, widely used to assess kidney impairments or complications related to diabetes~\cite{cite_040,cite_041}. The eGFR can be calculated using the modification of diet in renal disease (MDRD)  equation~\cite{cite_042,cite_043}, which considers CR, age, ethnicity, and gender, as follows:

\vspace{-.3cm}
\begin{align*}
\begin{split}
    eGFR = 186 \times (Cr)^{-1.154} \times (Age)^{-0.203} \\ \times \rho_{gender} \times \sigma_{ethnicity}
\end{split}
\tag{3}
\end{align*}
 
Here, $Cr$ signifies serum creatinine level measured in mg/dL. $Age$ refers to the individual's age in years. The term '$\rho_{gender}$' acts as a gender-based correction factor, assigning a value of $0.742$ for females and $1$ for males. '$\sigma_{ethnicity}$' is an ethnicity-based correction factor, with a value of $1.212$ for individuals of Black ethnicity and $1$ for those of other ethnicities.

The MDRD equation elucidates the causal relationships between eGFR and CR, and between Age and eGFR. In the MDRD equation, CR is used directly to estimate eGFR. Typically, an increase in CR suggests a decrease in eGFR, indicating reduced kidney efficiency~\cite{cite_dis09}. Moreover, age is a significant factor affecting eGFR. As age grows, their kidney function, as measured by eGFR, generally declines~\cite{cite_dis10}, establishing a causal relationship between increasing age and decreasing eGFR. Besides, the relationship between Gender and CR is causal, with men typically producing more CR due to higher muscle mass compared to women~\cite{cite_dis11}. These relationships are defined as causal, supported by the correlations in the MDRD equation and the absence of other potential common variables that could influence these changes in the literature.

\subsection{Bayesian Structural Time Series Forecasting}
Upon completing the construction of the Bayesian network, we employ a BSTS model to forecast future glucose values over various PHs. The BSTS model is a typical structure time series model, one based on the state space model, which represents time-series data as a combination of several components. The model’s decomposition capability aids in creating a modular and adaptable architecture, allowing for component customization in consideration of data characteristics. In this work, we incorporated various components such as trend, seasonal, and regressor components into the model. To further enhance performance, a spike-and-slab prior was applied to the regressor component. The BSTS model was then optimized using posterior distributions derived from multiple MCMC draws, and the final forecasting results were procured by averaging the forecastings from these optimized BSTS models. The definitions of the BSTS model components utilized in this project are as follows:

\begin{subequations}
\begin{align*}
    y_{t} &= \mu _{t} + \tau _{t} + \beta ^{t}X_{t} + \varepsilon _{t} \tag{4a}\\
       \mu_{t} &=  \mu_{t-1} + \delta _{t-1} + u_{t} ,\quad \delta _{t} = \delta _{t-1} + v_{t}     \tag{4b}\\ 
    \tau _{t} &= -\sum_{s=1}^{S-1}\tau _{t-s}+w_{t} \tag{4c} \\
\end{align*}
\end{subequations}

\vspace{-.5cm}
    
In these equations, $y_{t}$ represents the observed data at time $t$. $\mu_{t}$ and $\tau_{t}$ correspond to the trend and seasonal components at time $t$, respectively, $\mu_{t}$ is the sum of the level of trend at $\mu_{t-1}$, the slope of change of the trend indicated by $\delta_{t-1}$, and Gussian noise is represented by $u_{t}$ at previous time $t-1$. $\tau_{t}$ is the sum of the lagged values of $\tau _{t-s}$ up to $S-1$ lags, where $S$ signifies the seasonal period and $w_{t}$ denotes Gaussian noise. $\beta^{t}X_{t}$ refers to the regressor component at time $t$, with $\beta^{t}$ marking time-varying regression coefficients, while $X_{t}$ represents external regressors that influence the observed data. Furthermore, $\varepsilon_{t}$ is the error term at time $t$, which accounts for unpredicted variations in the observed data. $u_{t}$, $v_{t}$, and $w_{t}$ are the independent Gaussian random noises variables. In subsequent subsections, we detail the individual components decomposed from the BSTS model in our project.

\subsubsection{Semi-local Liner Trend Component}
The trend component represents the underlying long-term or systematic movement in the time series data. As presented in Formula 4b, this trend component captures the overall trajectory and sustained patterns in the data, which are not due to seasonal or cyclical fluctuations. Within the context of the BSTS model, we utilized a semi-local linear trend component to signify the long-term movements of glucose changes. The level and slope of the trend component are defined as follows:

\vspace{-.3cm}
\begin{align*}
\begin{split}
\mu_{t+1} &= \mu_t + \delta_t + \text{rnorm}(1, 0, \sigma_{\text{level}}) \\
\delta_{t+1} &= D + \phi \cdot (\delta_t - D) + \text{rnorm}(1, 0, \sigma_{\text{slope}})
\end{split}
\tag{5}
\end{align*}

In these equations, $\sigma_{\text{level}}$ and $\sigma_{\text{slope}}$ represent the inverse gamma priors on the level and slope standard deviations, respectively. $D$ refers to a Gaussian prior on the long-term slope parameter, while $\phi$ denotes the coefficient of a potential truncated Gaussian prior within the autoregressive (1) model, truncated within the range $(-1,1)$.

The semi-local linear trend assumes that similar to the local linear trend, the level component follows a random walk. However, the slope component is modeled to evolve according to an autoregressive (1) process, a basic type of time series model where future values are predicted based on a weighted sum of past observations, centered at a non-zero value, $D$. Therefore, the semi-local linear trend model ensures that the variance remains bounded over time, thereby providing more reliable uncertainty estimates for long-term forecasts. A more detailed analysis of the semi-local linear trend can be found in the references~\cite{cite_finaladd01,cite_finaladd02,cite_finaladd03}.

\subsubsection{Seasonal Component}

The seasonal component, as shown in Formula 4c, is used for capturing and modeling periodic patterns or seasonality inherent in time series data. Numerous factors (e.g., holidays, weather, etc.) occur periodically, thereby influencing the forecasted results. Thus, the integration of a seasonal component in the BSTS model enhances its capacity to separate long-term trends and seasonality from short-term fluctuations effectively.

The study focuses on the seasonal patterns in CGM values, mainly caused by daily routines such as general activities and food intake. Daily activities refer to daily life events and movements, inducing cyclical changes in CGM values. Food intake, typically three times a day, poses significant challenges for direct incorporation into forecast models due to the unquantified composition of food names and intake weights of raw data. To address this limitation, GL is introduced to quantify dietary influences, thus enhancing the model’s forecasting performance. Furthermore, these dietary events, which are not uniformly distributed within a 24-hour period of one day, exhibit a circadian pattern. According to the research~\cite{cite_finaladd04}, typical Chinese dietary events occur at approximately 7:00 am, 12:00 pm, and 6:00 pm, consistent with the data from both the ShanghaiT1DM and ShanghaiT2DM datasets. Thus, three distinct seasonal components were added to the model to effectively capture these seasonal patterns from the raw CGM data recorded every 15 minutes.

\textbf{Day Seasonal Component:} This component is divided into four seasons, representing four 15-minute intervals within an hour, and a duration of 24, signifying the 24 hours in a day. The configuration sets a day as a cyclical unit and captures daily variations in CGM values influenced by routine activities.

\textbf{Meal Seasonal Component:} Designed with a periodicity of 32 intervals and a frequency of three, this component divides the day into three 8-hour slots, corresponding to three times dietary events in one day. The objective of the meal seasonal component is to capture the influence of dietary events on CGM values. Nonetheless, as the dietary events do not distribute uniformly throughout the day, a single meal seasonal component can not accurately capture dietary influences.

\textbf{Circadian Seasonal Component:} The purpose of the circadian seasonal component is to differentiate between dietary influences and activities during daytime and nighttime, capturing their respective impacts on glucose levels. Since sleeping and active periods are asymmetric within a day, the circadian seasonal component is designed asymmetrically. Typically, humans are active for 16 hours and sleep continuously for 8 hours each day~\cite{cite_addT4}. Therefore, one circadian season consists of 48 intervals, representing 18 hours of activity per day, followed by 24 intervals representing 6 hours of sleep per day.

We validated their importance through ablation studies, which entailed removing each component sequentially and observing the effects on the model's performance. The outcomes of these studies are detailed in Section III-C.

\subsubsection{Regressor Component}
In this study, the BSTS model's regressor component was employed to capture the influence of external variables on the analyzed time series data. Specifically, CGM and GL values, extracted from two similar distributions selected via consensus Bayesian network, were incorporated as regressor components. The goal is to consider similar subjects' CGM data and GL values as external factors impacting the behavior of time series forecasting. To discern the most relevant regression components affecting the CGM values, a statistical technique known as spike-and-slab prior was employed. This technique aids in identifying the specific element within the CGM data that significantly contributes to variability within the time series while effectively filtering out irrelevant or noise-induced components. By leveraging the regressor component and this statistical technique, this study aims to enhance modeling and forecasting performance for the time series data, particularly when dealing with intricate relationships and multiple variables influencing the time series dynamics. A series of ablation experiments, elaborated upon in Section II-D, further elucidate these similar subjects' distribution's impact on the model's performance.

\subsubsection{Average Forecasting and MCMC}

When dealing with models that have intricate likelihood structures or that are highly complex, obtaining closed-form solutions for the posterior distributions often becomes challenging or even infeasible. The MCMC method provides an efficient approach to approximate the posterior distribution through an iterative sampling process. In this study, where Bayesian model averaging is integrated into the BSTS model, we deploy MCMC with 1000 iterative draws to approximate the posterior model probabilities that are sampled from the joint posterior distribution of model parameters. The final step involves model averaging to get the final forecasts, taking into account uncertainties associated with both the model parameters and the model itself. Model average forecasting, combined with MCMC, ensures a robust and comprehensive forecast. It effectively accommodates various sources of uncertainty and variability inherent in the modeling process, while also preventing over-fitting by minimizing reliance on any specific dataset characteristics.

\section{Experiments and Results}

\subsection{Experiments Settings}

\begin{table*}[htb!]
\centering
\begin{tabular}{lcclcccc}
\toprule
\multirow{2}{*}{Methods} &\multirow{2}{*}{Datasets}&\multirow{2}{*}{Forecasting models}& \multirow{2}{*}{Metrics} & \multicolumn{4}{c}{Prediction horizons} \\
\cline{5-8}
& &&& 15-min & 30-min & 45-min & 60-min \\ \midrule

\multirow{2}{*}{Li et al.~\cite{cite_intro06}} & \multirow{2}{*}{UVA/Padova} &\multirow{2}{*}{CNN} & RMSE (mg/dL) & N/A & $10.73\pm1.66$ & N/A & $22.65\pm4.71$ \\
& && MAPE (\%) & N/A & $5.77\pm0.88$ & N/A & $11.23\pm1.96$ \\
\cmidrule{1-8}

\multirow{2}{*}{Deng et al.~\cite{cite_res01}} & \multirow{2}{*}{OhioT1DM} &\multirow{2}{*}{CNN} & MAE (mg/dL) & N/A & $13.53$ & N/A & $24.65$ \\
&& & RMSE (mg/dL) & N/A & $19.08$ & N/A & $33.80$ \\
\cmidrule{1-8}

\multirow{3}{*}{Zhu et al.~\cite{cite_res02}} & \multirow{3}{*}{OhioT1DM} &\multirow{3}{*}{RNN} & MAE (mg/dL) & $7.21\pm1.09$ & $15.06\pm2.36$ & $21.15\pm3.15$ & $26.11\pm4.36$ \\
& && RMSE (mg/dL) & $10.15\pm1.67$ & $20.92\pm3.55$ & $28.99\pm4.41$ & $35.28\pm5.77$ \\
& && MAPE (\%) & $5.07\pm0.97$ & $10.62\pm2.03$ & $14.94\pm2.77$ & $18.53\pm3.78$ \\
\cmidrule{1-8}

\multirow{3}{*}{Zhu et al.~\cite{cite_addT3}} & \multirow{3}{*}{ShanghaiDM} &\multirow{3}{*}{Transformer} & MAE (mg/dL) &  N/A& $8.8\pm2.8$ & N/A  & $15.1\pm5.1$ \\
& && RMSE (mg/dL) &  N/A & $12.7\pm3.8$ &  N/A & $21.7\pm6.9$ \\
& && MAPE (\%) &  N/A & $6.7\pm2.3$ & N/A & $11.2\pm4.1$ \\
\cmidrule{1-8}

\multirow{3}{*}{ARIMAX~\cite{cite_res03}} & \multirow{3}{*}{ShanghaiT2DM} &\multirow{3}{*}{ARIMAX}& MAE (mg/dL) & $8.59 \pm 0.92$ & $13.83\pm 1.84$ & $19.98\pm 2.81$ & $20.43\pm 4.19$ \\
&& & RMSE (mg/dL) & $11.91\pm 1.41$ & $20.14\pm 3.54$ & $27.18\pm 4.67$ & $29.41\pm 6.03$ \\
& && MAPE (\%) & $7.39\pm 0.48$ & $11.61\pm 1.37$ & $16.85\pm 2.09$ & $18.92\pm 3.41$ \\
\cmidrule{1-8}
\multirow{3}{*}{Ours} & \multirow{3}{*}{ShanghaiT2DM} &\multirow{3}{*}{BSTS}& MAE (mg/dL) & $6.41 \pm 0.60$ & $8.91\pm 1.21$ & $11.74\pm1.85$ & $14.94\pm2.81$ \\
& && RMSE (mg/dL) & $8.29 \pm 0.95$ & $13.52\pm 2.42$ & $17.01\pm3.03$& $21.41\pm 4.12$ \\
& && MAPE (\%) & $5.28\pm 0.33$ & $7.02\pm 0.94$ & $9.09\pm1.27$ & $11.68\pm 2.57$ \\

\bottomrule
\end{tabular}
\caption{Comparison results (Mean$\pm$SD) of error metrics (MAE, RMSE, MAPE) for BSTS  and ARIMAX models of ShanghaiT2DM dataset and other state-of-the-art methods applied to the UVA/Padova simulator and OhioT1DM datasets.}
\label{tab:my-table_4}
\end{table*}

\begin{figure*}[htb!]
  \centering
  \centerline{\includegraphics[width = 1\linewidth,height = 3in]{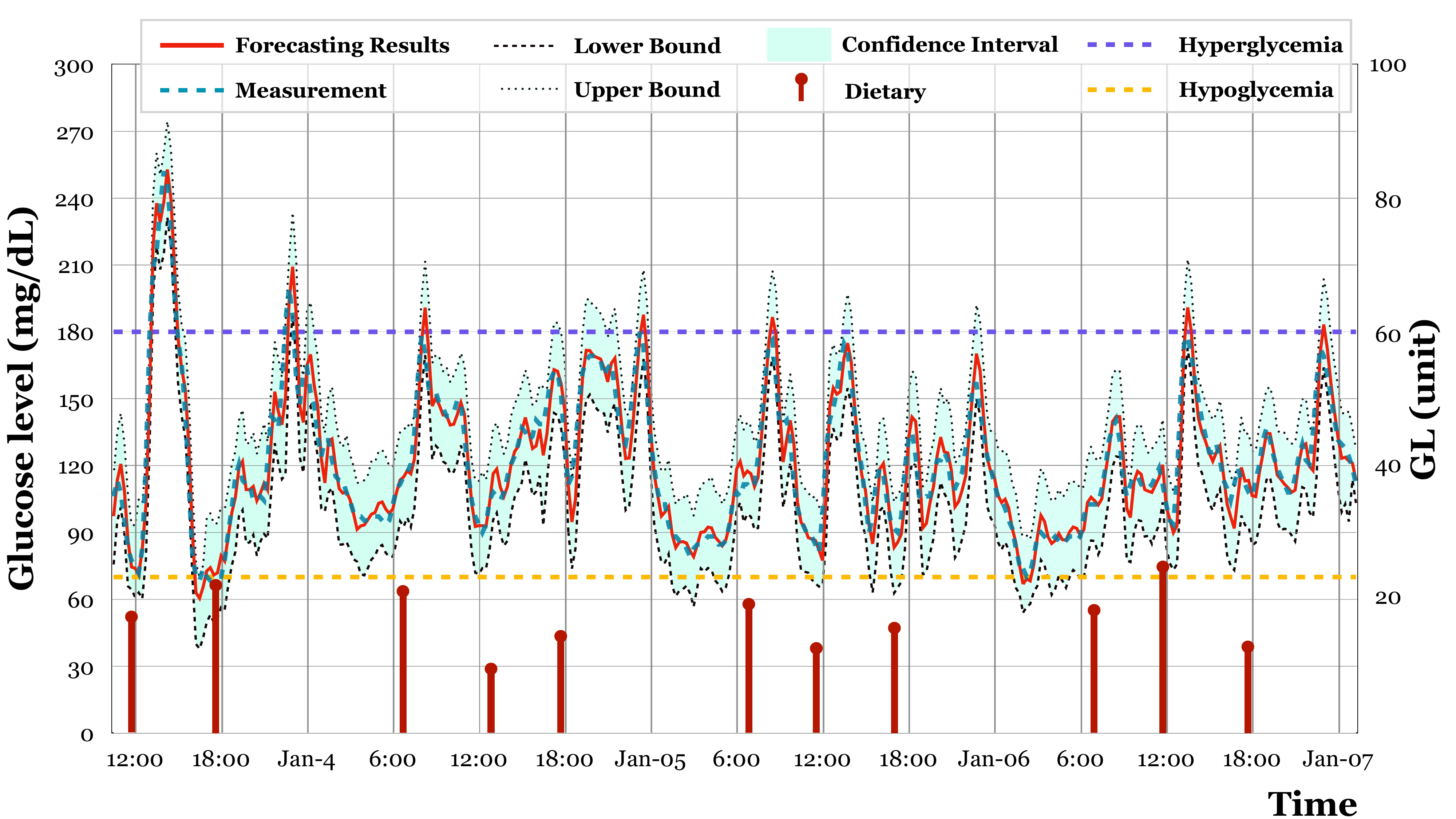}}
  \caption{Three-day period of CGM and forecasting trajectories for a Shanghai T2DM subject over a 15-minute prediction horizon, including 95\% confidence interval upper and lower bounds of forecasts.}
\label{fig:fig_label_01}
\end{figure*}

We conducted two series of experiments to evaluate our proposed model: the first series involved comparative experiments to benchmark against other state-of-the-art (SOTA) works, and the second series included ablation studies to validate the components of the BSTS model. In both experiments, we first divided the time-series data using an 80:20 split ratio. For each subject, we allocated the first 80\% of the data to the training set and the remaining 20\% to the testing set, with the last 20\% of the training set designated as the validation set. The training set was used for model development, the validation set for preventing over-fitting, and the final model performance was evaluated on the testing set. This data split is widely adopted in existing research on SOTA blood glucose forecasting methods~\cite{cite_intro06,cite_res01,cite_res02,cite_addT3}. Additionally, since the processed time-series data spans durations ranging from 3 to 14 days, allocating 80\% to the training set ensures the inclusion of at least a 2-day period of data. This allocation is crucial for effectively capturing both trend and seasonal components during model training, particularly the daily seasonal component. Activating the daily seasonal component helps capture cyclic daily variations, thereby enhancing the model's forecasting performance, as demonstrated by the ablation analysis results listed in Table V.

During the evaluation test, we utilized a sliding window mechanism designed to span 8 intervals, equivalent to a 2-hour period of time series data. Our objective was to use the data within the sliding window to forecast blood glucose values through 1000 MCMC draws of the BSTS model for subsequent PHs of $15$-minute, $30$-minute, $45$-minute, and $60$-minute, and to evaluate their respective performances. During testing, only the test subject's data within the testing set and the designated sliding window size are incorporated, rather than data spanning longer periods used for training. The training, validation, and testing sets are completely independent of each other, and no testing data is introduced during model training. Notably, the forecasting experiments were conducted exclusively on the ShanghaiT2DM dataset due to the limited size of the ShanghaiT1DM dataset.

In these experiments, the evaluation criteria consist of the mean values of metrics like mean absolute error (MAE), root mean square error (RMSE), and mean absolute percentage error (MAPE), and their corresponding SD. The units of both MAE and RMSE are standardized as mg/dL, and the unit of MAPE is \%, aligning with common evaluation criteria for CGM forecasting models. The formula 6 for MAPE calculation is as follows:

\begin{align*}
    MAPE=\frac{1}{n}\sum_{i=1}^{n}{\frac{\left | x_i - y_i\right|}{y_i}}*100\% \tag{6}
\end{align*}

Here, $x$ and $y$ are n-dimensional vectors, where $x$ pertains to the CGM values recorded in the time-series dataset, $y$ signifies the forecasting output values from the BSTS model, and $n$ denotes the total count of samples in the test set.

\subsection{Evaluation of Forecasting Performance on the ShanghaiT2DM Dataset: Results Across Various Prediction Horizons and Different Models}

In this study, we evaluated the BSTS model's forecasting performance in comparison to the autoregressive integrated moving average with explanatory variable (ARIMAX) model~\cite{cite_res03} as a benchmark. These experiments were conducted on the ShanghaiT2DM dataset across multiple PHs of 15, 30, 45, and 60 minutes. The ARIMAX model, an extension of the ARIMA model, incorporates external predictor variables, reducing forecasting errors. The experimental configurations for each model are detailed as follows:

\textbf{BSTS}: In our experiments, we selected a single subject's samples for testing. Utilizing the Bayesian network, we identified two subjects with similar distributions, and we integrated GL records and CGM values from these similar subjects and test subjects into the BSTS model as described in the Method section. We then employed a growing window approach~\cite{cite_044} to forecast the CGM values across the specified PH.

\textbf{ARIMAX}: The classical ARIMAX model with fixed parameters for forecasting the selected subject's CGM values, which were set as labels. External variables included the test subject's GL records, along with CGM values and GL records from two similar subjects identified via the Bayesian network. The training and testing procedures were the same as those applied in the BSTS experiments. The ARIMAX model parameters ($p$, $d$, $q$) were fixed at $(3, 0, 2)$ following a selection process. An augmented Dickey-Fuller test confirmed no differencing was needed for stationarity ($d=0$). Parameters $p$ and $q$ were identified based on autocorrelation and partial autocorrelation function plots. A grid search was then conducted to optimize these parameters by minimizing the Akaike information criterion.


Table \ref{tab:my-table_4} lists the results of experiments, including the comparison experiments performance of ARIMAX and BSTS models on the ShanghaiT2DM dataset against other SOTA methods~\cite{cite_intro06,cite_res01,cite_res02} on the UVA/Padova T1DM simulator and the OhioT1DM dataset. Some entries in Table~\ref{tab:my-table_4} are noted as 'N/A' or lack standard deviations (SD), denoting that specific PHs were not evaluated in their original studies.

\begin{figure*}[htb!]
  \centering
  \centerline{\includegraphics[width = 1\linewidth,height = 3in]{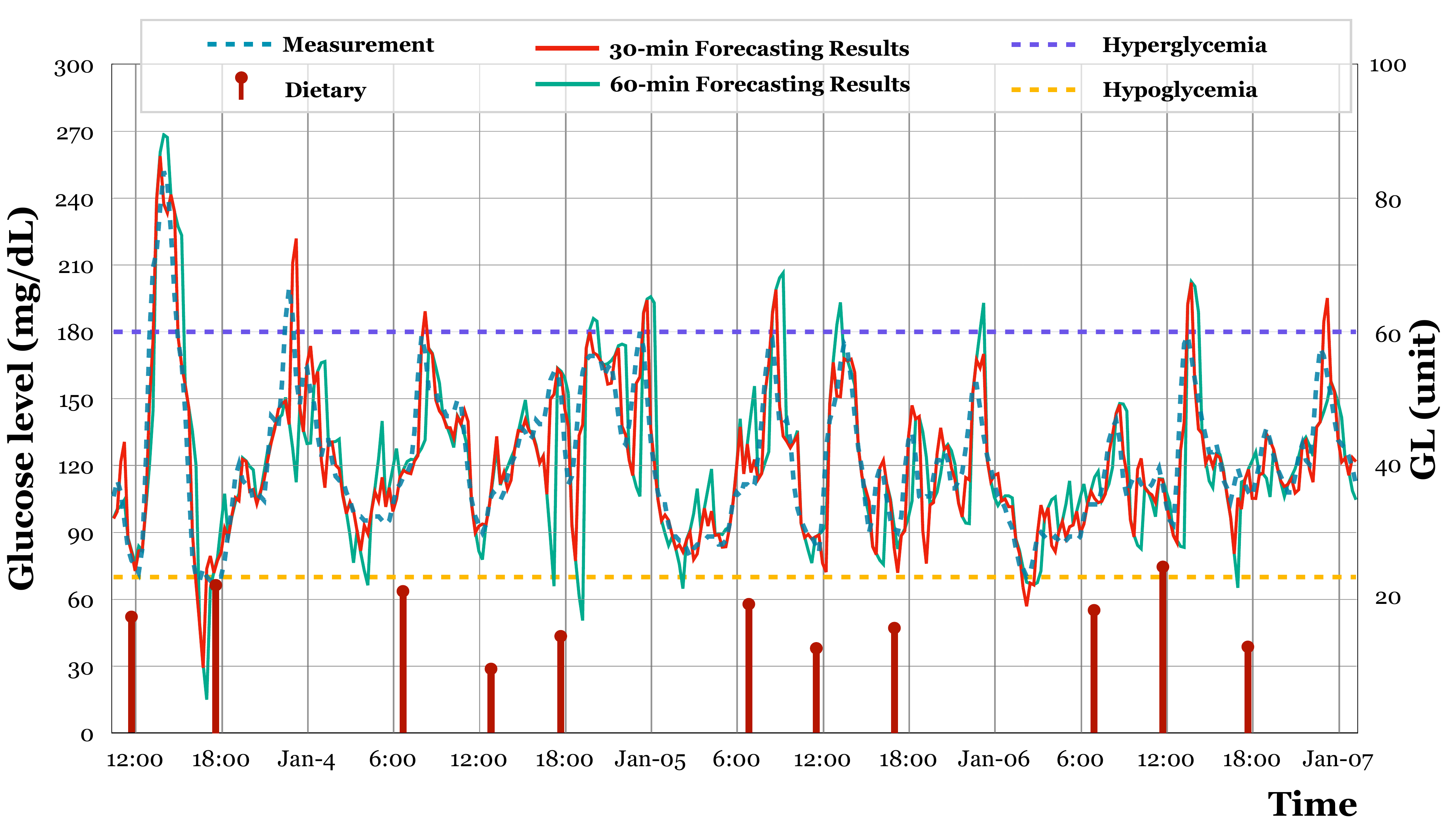}}
  \caption{Three-day period of CGM and forecasting trajectories for a Shanghai T2DM subject over the 30-minute and 60-minute prediction horizon.}
\label{fig:fig_label_03}
\end{figure*}

\begin{figure*}[htb!]
  \centering
  \centerline{\includegraphics[width = .9\linewidth,height = 1.7in]{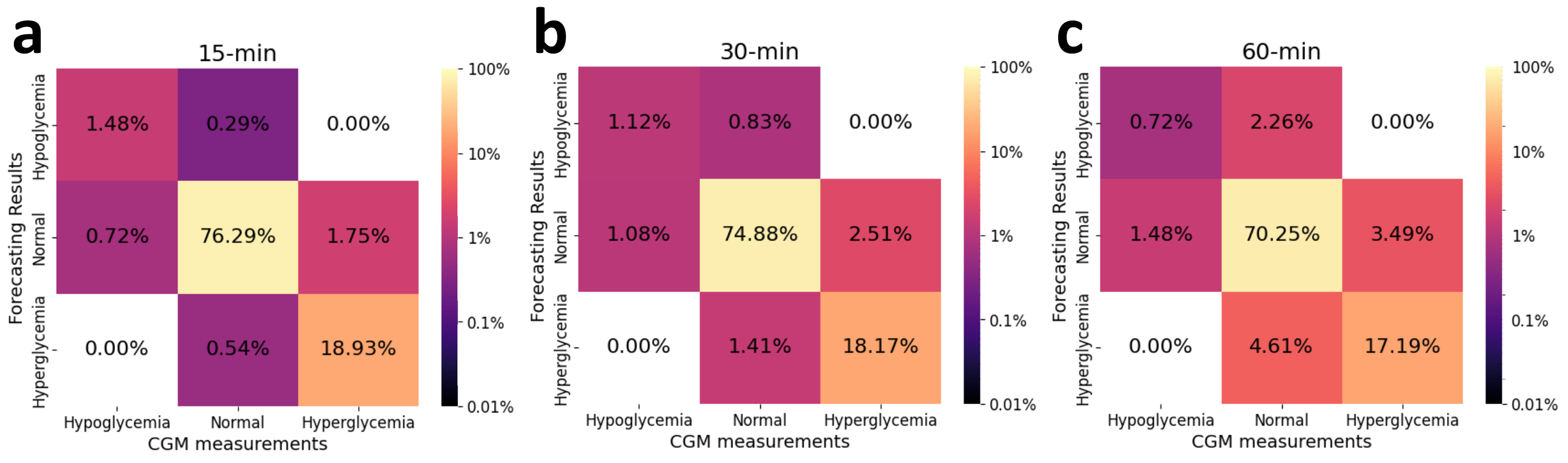}}
  \caption{Confusion matrices for blood glucose forecasting results at (a) 15-minute, (b) 30-minute, and (c) 60-minute prediction horizons. Each matrix compares predicted glucose conditions (hypoglycemia, normal, hyperglycemia) against actual CGM measurements. The percentages indicate the proportion of all subjects' forecasting results falling into each category.}
\label{fig:fig_label_04}
\end{figure*}

Li et al.~\cite{cite_intro06} evaluated their customized convolutional neural network (CNN)-based model on 10 T1DM subjects on the UVA/Padova simulator. Their forecasting results for the 30-minute PH showed a RMSE of $10.73\pm1.66$ and a MAPE of $5.77\pm0.88$. While no results were available for the 15 and 45-minute PHs, the results of the 60-minute forecast had a RMSE of $22.65\pm4.71$ and a MAPE of $11.23\pm1.96$. Deng et al.~\cite{cite_res01} and Zhu et al.~\cite{cite_res02} both applied their methods to the OhioT1DM dataset. Zhu et al.~\cite{cite_res02} incorporated all 12 subjects into their recurrent neural network (RNN)-based model, whereas Deng et al.~\cite{cite_res01} limited their experiments to 6 subjects using a CNN-based transfer model. In the comparison, Deng et al.~\cite{cite_res01} outperformed Zhu et al.~\cite{cite_res02} at both the 30-minute and 60-minute PHs.

In our study, we compared our proposed method's performance with the ARIMAX baseline model, utilizing the ShanghaiT2DM dataset comprising 91 samples from 83 subjects. Since the ARIMAX model utilized fixed parameters without recursive updates during the training and testing phases, both the input and lagged data remained constant. As a result, our method consistently outperformed the ARIMAX model, particularly at extended PHs of 45 and 60 minutes, demonstrating enhanced robustness and fewer errors. Moreover, our model demonstrated superior forecasting performance across various PHs compared to other SOTA methods, evaluated on a larger and more diverse dataset. Notably, when compared to the transformer-based forecasting model used by Zhu et al.~\cite{cite_addT3}, which was applied to the same dataset without differentiating between types of diabetes mellitus, our model displayed comparable performance with significantly reduced standard deviation. This highlights the robustness and adaptability of our method in processing extensive, heterogeneous data.

Figure \ref{fig:fig_label_01} visualizes the forecasting CGM trajectories through the BSTS method and the actual CGM measurements of a single subject. In this experiment round, we utilized an initial two-day training period and a three-day forecasting window. The quantified representation of daily dietary records as GL values in Figure \ref{fig:fig_label_01} shows the potential impact of dietary events on the fluctuations in CGM measurements. Additionally, the figure includes upper and lower bounds (95\% confidence interval) of the forecasts, demonstrating the robustness and reliability of our forecasting model.

\begin{table*}[htb!]
\centering
\begin{tabular}{@{}llcccc@{}}
\toprule
\multirow{2}{*}{Ablation experiments} & \multirow{2}{*}{Metrics} & \multicolumn{4}{c}{Prediction horizons} \\
\cline{3-6}
& & 15-min & 30-min & 45-min & 60-min \\ \midrule

\multirow{3}{*}{Removing similar subjects' distribution} & MAE (mg/dL) & $11.12 \pm 0.98$ & $15.67 \pm 2.21$ & $19.03 \pm 2.80$ & $25.59 \pm 3.44$ \\
& RMSE (mg/dL) & $13.74 \pm 1.34$ & $21.88 \pm 2.36$& $25.73 \pm 3.28$ & $33.45 \pm 4.27$\\
& MAPE (\%) & $9.20 \pm 0.55$ & $13.11 \pm 1.27$ & $15.28 \pm 1.79$ & $21.46 \pm 2.68$ \\ \midrule

\multirow{3}{*}{Removing day seasonal component}& MAE (mg/dL) & $10.57 \pm 1.09$ & $16.80 \pm 2.16$ & $20.56 \pm 2.67$& $26.87 \pm 3.45$ \\
& RMSE (mg/dL) & $13.07 \pm 1.30$ & $22.09 \pm 2.47$ & $27.94 \pm 3.20$ & $35.65 \pm 4.48$ \\
& MAPE (\%) & $9.45 \pm 0.56$ & $13.43 \pm 1.58$ &$18.69 \pm 1.99$ & $22.82 \pm 2.81$ \\ \midrule

\multirow{3}{*}{Removing meal seasonal component}& MAE (mg/dL) & $8.91 \pm 0.91$ & $14.74 \pm 1.98$ & $19.56 \pm 3.27$ & $23.25 \pm 3.02$ \\
& RMSE (mg/dL) & $11.54 \pm 1.09$ & $19.95 \pm 2.26$ & $27.88 \pm 3.54$ & $30.27 \pm 3.60$ \\
& MAPE (\%) & $8.27 \pm 0.51$ & $11.70 \pm 1.17$ & $16.58 \pm 2.12$ & $17.84 \pm 2.74$ \\ \midrule

\multirow{3}{*}{Removing circadian seasonal component}& MAE (mg/dL) & $8.84 \pm 0.88$ & $14.34 \pm 1.95$ & $19.01 \pm 2.67$ & $20.02 \pm 3.08$ \\
& RMSE (mg/dL) & $12.56 \pm 1.01$ & $20.05 \pm 2.29$ & $26.96 \pm 3.14$ & $28.18 \pm 3.79$ \\
& MAPE (\%) & $7.63 \pm 0.47$ & $10.71 \pm 1.13$ & $15.79 \pm 1.84$ & $16.59 \pm 2.53$ \\ \midrule

\multirow{3}{*}{Baseline} & MAE (mg/dL) & $6.58 \pm 0.59$ & $8.92\pm 1.37$ & $11.79\pm1.87$ & $15.01\pm2.65$ \\
& RMSE (mg/dL) & $8.77\pm 0.99$ & $13.59\pm 2.25$ & $17.52\pm3.04$ & $21.06\pm 3.97$ \\
& MAPE (\%) & $5.34\pm 0.38$ & $7.53\pm 1.01$ & $9.89\pm1.26$ & $11.62\pm 2.32$ \\ \bottomrule

\end{tabular}
\caption{Performance (Mean$\pm$SD) of ablation experiments on the BSTS model. The baseline refers to the BSTS model's results using the same setting as the comparison experiments described in Section IV-B.
}
\label{tab:my-table_5}
\end{table*}

Figure \ref{fig:fig_label_03} shows the forecasting CGM trajectories for the same subject over 30-minute and 60-minute PH. In comparison to the 15-minute PH forecasts, errors are more noticeable in these longer PHs, particularly at inflection points where the forecasted values change more drastically. This finding corresponds to confusion matrices shown in Figure \ref{fig:fig_label_04}, where the accuracy of classifying hypoglycemia, normal, and hyperglycemia decreases notably as the PH increases. Specifically, the overall accuracy of all subjects decreases from 96.70\%, 94.17\%, to 88.16\% as the PH increases. The trend is a decrease in classification accuracy across all categories, with an increasing number of measurements that were originally normal being incorrectly forecasted as hyperglycemia at extended PHs. However, these incorrect classifications of hyperglycemia are not the primary sources of forecasting errors. As shown in Table \ref{tab:my-table_9}, the range of forecasted glucose levels increases with the PH, especially in hypoglycemia areas. Forecasts with large errors occur in hypoglycemia areas rather than in hyperglycemia areas when compared to the true measurement range.

\begin{table}[htb!]
\centering
\begin{tabular}{ccc}
\toprule
Prediction & Measurement   &  Forecasting  \\ 
horizons & ranges (mg/dL) & ranges(mg/dL) \\
\midrule
15-minute&(39.6, 468.0)&(24.33, 470.53)\\
30-minute&(39.6, 468.0)&(16.47, 476.69)\\
60-minute&(39.6, 468.0)&(9.86, 485.74)\\
\bottomrule
\end{tabular}
\caption{Ranges (minimum, maximum) of glucose measurements and corresponding predicted values for 15, 30, and 60-minute prediction horizons.}
\label{tab:my-table_9}
\end{table}

\subsection{Ablation Experiments of the BSTS Model Components}

We conducted ablation experiments to evaluate the impact of specific components within the BSTS forecasting model. Ablation studies, by systematically removing individual components, provide a robust method to observe their influence on model performance. We designed four distinct ablation scenarios: one involving the removal of similar subject distributions as determined by the Bayesian network, and three others separately removing each of the seasonal components within the BSTS model. These three seasonal components including day, meal, and circadian within this model allow the capture of daily activities and dietary patterns' effects on CGM values. 

The comparison experiments, as detailed in Section III-B, were conducted 91 times for one PH forecasting, while each ablation experiment was performed a limited 20 times. For the comparison experiment, every sample was used as a unique test instance in each PH forecasting experiment. In contrast, the ablation experiments randomly selected subjects 20 times, with the average results and their corresponding SD listed in Table~\ref{tab:my-table_5}. Except for these changes, all other experimental settings remained consistent with those described in Section III-B.

Through the analysis of ablation experiments, we observed that when we removed the similar subjects' CGM trajectories and GL records, sourced from the Bayesian network, there was a marked increase in error across all metrics. The increased error rates underscore the critical role of incorporating similar subjects' data for accurate glucose forecasting. The model's sensitivity to this ablation experiment effectively suggests individual physiological differences significantly influencing blood glucose changes and introducing other similar individuals' information can reduce these influences. Therefore, it is necessary to build personalized blood glucose forecasting models to achieve accurate forecasting results.

Furthermore, an ablation experiment was conducted with a 30-minute PH to determine the optimal number $M$ of similar individuals in Algorithm 1. The results of this experiment, shown in Table \ref{tab:my-table_7}, evaluated various settings for $M$ in Algorithm 1, ranging from 1 to 4. Due to dataset size constraints, the maximum tested value for $M$ was limited to 4.

\begin{table}[htb!]
\centering
\begin{tabular}{cccc}
\toprule
M&  MAE (mg/dL) & RMSE (mg/dL) & MAPE (\%)  \\ \midrule

1 & $12.19 \pm 1.73$ & $17.14 \pm 2.32$ & $10.79 \pm 1.06$ \\
2 & $8.65 \pm 1.08$ & $13.28 \pm 2.11$ & $6.90 \pm 0.97$ \\
3 & $11.75 \pm 1.66$ & $16.51 \pm 2.50$ & $9.54 \pm 1.28$ \\ 
4 & $13.84 \pm 1.78$ & $18.19 \pm 3.07$ & $13.03 \pm 1.75$ \\ 

\bottomrule
\end{tabular}
\caption{Performance (Mean$\pm$SD) of ablation experiments on the similar individuals introduced on the BSTS model (PH = 30-minute). M refers to the number of similar individuals selected in Algorithm 1.
}
\label{tab:my-table_7}
\end{table}

The results in Table \ref{tab:my-table_7} demonstrate that the model achieves its best performance when $M$ is set to 2. While the number $M$ of similar individuals included in the model is a critical variable, other factors, such as the causes within the ShanghaiT2DM dataset (insulin sensitivity and glucose sensitivity), are able to influence the results. As $M$ increases, more subject information is introduced, potentially adding more variance related to T2DM causes into the model, which has the potential to affect performance. However, the impact of these T2DM causes is minimal, controlled, and further discussed in the comprehensive analysis provided in the Discussion section.

\section{Discussion}
Our approach is characterized by the integration of data-driven and knowledge-driven approaches for analysis. The Bayesian network’s arc strength is crucial in determining the existence and strength of connections within the network, thereby influencing the inferred variables FPG and 2HPP. The arc strength directly impacts the process of selecting similar individuals within our study by affecting the generated structure and finally affecting the precision of our model. The Types of arcs in our Bayesian network are data-driven classifications, including consensus, unique, and possible arcs, which are derived based on their presence in Model\_1 and Model\_2 and their respective arc strengths. The categorization of arcs into causal, correlated, and independent relationships is by a comprehensive knowledge-driven analysis, which enhances the clinical reliability of our findings. Importantly, all arcs categorized as causal or correlated relationships correspond to either consensus or unique arc types, while those categorized as independent relationships align with possible arc types. This categorization shows that all consensus and unique arcs generated through the data-driven process are validated by knowledge-driven analysis and have clinical correspondence, thereby significantly enhancing the interpretability of our model.

In our comparison experiments, the proposed method demonstrated superior performance over the ARIMAX baseline model and other SOTA  methods. Particularly at extended prediction horizons of 45 and 60 minutes, our method demonstrated enhanced robustness and fewer errors. This is a significant result, considering the complex, heterogeneous, and large size of the dataset involved in ShanghaiT2DM research. Compared to SOTA methods applied to other public CGM datasets, our model can effectively process and analyze such data suggesting its potential applicability in diverse clinical individuals.

The ablation experiments on the BSTS model's seasonal components provide insights into the multifaceted factors influencing blood glucose levels. The removal of day seasonal component led to a notable increase in forecast errors, particularly at longer PHs. This highlights the role of daily seasonal patterns in blood glucose regulation, showing the model's reliance on accurately capturing these 24-hour seasons. The relationship between dietary activities and blood glucose levels is complex, influenced by factors like food intake, meal timing, and individual metabolic responses. The exclusion of meal seasonality revealed the direct association between dietary events and subsequent glucose level fluctuations, particularly at longer PHs. This finding agrees with related research indicating that postprandial glucose peaks often occur at longer intervals after meals, with average peaks appearing around 72$\pm$23 minutes postprandial~\cite{cite_disadd01}. The circadian rhythm, reflecting day-night changes, significantly influences blood glucose levels. A marked decrease in forecasting errors was observed upon removing the circadian seasonal component. This result indicates the circadian pattern's effect on glucose forecasting, reflecting daily activities and physiological variations between day and night, which is crucial in glucose level fluctuations. These ablation experiment results collectively affirm the importance of considering these three seasonal components in glucose level forecasting models.

However, the study faces limitations. The primary limitation lies in the dataset size. While the datasets for ShanghaiT2DM and ShanghaiT1DM are relatively large compared to other available glucose monitoring datasets, they are still limited in their capacity to develop a fully generalized model that can effectively capture the wide range of individual differences in T2DM. Future research should aim to expand these datasets in terms of size and diversity, enhancing the model's robustness and generalizability. Further, the integration of domain generalization algorithms is essential to address individual variations more effectively, a common challenge in personalized medicine. Such enhancements are crucial for the model to be applicable in a broader clinical context and to ensure that it can accommodate the complexities inherent in individual patient profiles.

Moreover, further analysis is required on how different causes of T2DM, such as insulin sensitivity and beta-cell glucose sensitivity, impact glucose tolerance and CGM fluctuations in subjects. Bergman et al.~\cite{cite_addT2} conducted a study using intravenous glucose tolerance tests (IVGTT) and two kinetic models to measure insulin and glucose sensitivity. Their results indicate that T2DM causes are associated with subjects' body weight, specifically, all cases of insulin sensitivity were observed in obese subjects. Based on these findings, we can preliminarily analyze the causes in the ShanghaiT2DM dataset, identifying obese subjects (those with an ideal body weight $>$ 130\%) and lean subjects. Obese subjects, potentially exhibiting insulin sensitivity, comprise only 3.3\% (3 out of 91) of all T2DM subjects in the dataset. Consequently, as the number $M$ of similar individuals increases, the potential to introduce subjects with insulin sensitivity is limited and infrequent, suggesting that variance in T2DM causes is not the primary influencer of experiment results. 

However, the daily scenarios of ShanghaiT2DM dataset collected are more complex than those in controlled IVGTT settings. The introduction of various dietary events can complicate the construction of dynamic models as previously done in~\cite{cite_addT2}. Therefore, future studies should aim to apply classic insulin and glucose kinetics models after obtaining more comprehensive records on the causes of T2DM and incorporating quantified dietary intakes and insulin injections. This approach will facilitate a deeper exploration of the dynamics of glucose tolerance for different T2DM causes in daily life.

In conclusion, we introduced a novel approach to glucose level forecasting in T2DM, enhancing our understanding of the disease’s complexities. By processing data, quantifying meal events as glucose load features, and creating three seasonal components, our model captures the diverse patterns influencing glucose fluctuations in daily life and provides accurate forecasts across various PHs. The integration of data-driven and knowledge-driven approaches in our methodology improves the interpretability of the model and ensures its reliability. This combination allows for a detailed understanding of T2DM, contributing to the development of more personalized and effective management strategies.

\section{Conclusion}
In this study, we have developed a novel blood glucose forecasting system specifically for the ShanghaiT2DM dataset, which integrates Bayesian network structure learning with the BSTS model. Our system forecasts glucose levels over intervals ranging from 15 to 60 minutes with low prediction errors, considering past CGM trajectories, dietary records, and individual differences. Table \ref{tab:my-table_4} presents the performance of our forecasting model across all evaluation metrics for these PHs. Specifically, for a 15-minute PH, the model shows a MAE of \(6.41 \pm 0.60\) mg/dL, a RMSE of \(8.29 \pm 0.95\) mg/dL, and a MAPE of \(5.28 \pm 0.33\%\). Comprehensive ablation studies have confirmed the effectiveness and components design of our BSTS forecasting approach. Moreover, our research on the Bayesian network structure illuminates the complex relationships among diabetes-related variables, setting a foundation for future research. With our work on the ShanghaiT2DM dataset, we contribute significantly to the field of T2DM research, offering a benchmark for subsequent studies and enhancing understanding in this area.

\end{document}